# SynthMorph: Learning Contrast-Invariant Registration Without Acquired Images

Malte Hoffmann<sup>ID</sup>, Benjamin Billot<sup>ID</sup>, Douglas N. Greve<sup>ID</sup>, Juan Eugenio Iglesias<sup>ID</sup>, Bruce Fischl<sup>ID</sup>, and Adrian V. Dalca<sup>ID</sup>

**Abstract**—We introduce a strategy for learning image registration without acquired imaging data, producing powerful networks agnostic to contrast introduced by magnetic resonance imaging (MRI). While classical registration methods accurately estimate the spatial correspondence between images, they solve an optimization problem for every new image pair. Learning-based techniques are fast at test time but limited to registering images with contrasts and geometric content similar to those seen during training. We propose to remove this dependency on training data by leveraging a generative strategy for diverse synthetic label maps and images that exposes networks to a wide range of variability, forcing them to learn more invariant features. This approach results in powerful networks that accurately generalize to a broad array of MRI contrasts. We present extensive experiments with a focus on 3D neuroimaging, showing that this strategy enables robust and accurate registration of arbitrary MRI contrasts even if the target contrast is not seen by the networks during training. We demonstrate registration accuracy surpassing the state of the art both within and across contrasts, using a single model. Critically, training on arbitrary shapes synthesized from noise distributions results in competitive performance, removing the dependency on acquired data of any kind. Additionally, since anatomical label maps are often available for the anatomy of interest, we show that synthesizing images from these dramatically boosts performance, while still avoiding the need for real intensity images. Our code is available at https://w3id.org/synthmorph.

Manuscript received July 28, 2021; revised September 20, 2021; accepted September 28, 2021. Date of publication September 29, 2021; date of current version March 2, 2022. This work was supported in part by Alzheimer's Research UK under Grant ARUK-IRG2019A-003; in part by the European Research Council (ERC) under Starting Grant 677697; and in part by the National Institutes of Health (NIH) under Grant 1R01 AG070988-01, BRAIN Initiative Grant 1RF1MH123195-01, Grant K99 HD101553, Grant U01 AG052564, Grant R56 AG064027, Grant R01 AG064027, Grant R01 AG016495, Grant U01 MH117023, Grant P41 EB015896, Grant R01 EB023281, Grant R01 EB019956, Grant R01 NS0525851, Grant R21 NS072652, Grant R01 NS083534, Grant U01 NS086625, Grant U24 NS10059103, Grant R01 NS105820, Grant S10 RR023401, Grant S10 RR019307, and Grant S10 RR023043. *(Corresponding author: Malte Hoffmann.)*

This work involved human subjects or animals in its research. Approval of all ethical and experimental procedures and protocols was granted by IRBs at Washington University in St. Louis (201603117) and Mass General Brigham (2016p001689).

Malte Hoffmann and Douglas N. Greve are with the Athinoula A. Martinos Center for Biomedical Imaging, Massachusetts General Hospital, Charlestown, MA 02129 USA, and also with the Department of Radiology, Harvard Medical School, Boston, MA 02115 USA (e-mail: mhoffmann@mgh.harvard.edu; dgreve@mgh.harvard.edu).

Benjamin Billot is with the Centre for Medical Image Computing, University College London, London WC1E 6BT, U.K. (e-mail: benjamin.billot.18@ucl.ac.uk).

Juan Eugenio Iglesias is with the Athinoula A. Martinos Center for Biomedical Imaging, Massachusetts General Hospital, Charlestown, MA 02129 USA, also with the Department of Radiology, Harvard Medical School, Boston, MA 02115 USA, also with the Centre for Medical Image Computing, University College London, London WC1E 6BT, U.K., and also with the Computer Science and Artificial Intelligence Laboratory, MIT, Cambridge, MA 02139 USA (e-mail: e.iglesias@ucl.ac.uk).

Bruce Fischl and Adrian V. Dalca are with the Athinoula A. Martinos Center for Biomedical Imaging, Massachusetts General Hospital, Charlestown, MA 02129 USA, also with the Department of Radiology, Harvard Medical School, Boston, MA 02115 USA, and also with the Computer Science and Artificial Intelligence Laboratory, MIT, Cambridge, MA 02139 USA (e-mail: bfischl@mgh.harvard.edu; adalca@mit.edu).

Digital Object Identifier 10.1109/TMI.2021.3116879

## I. INTRODUCTION

IMAGE registration estimates spatial correspondences between image pairs and is a fundamental component of many neuroimaging pipelines involving data acquired across time, subjects, and modalities. Magnetic resonance imaging (MRI) uses pulse sequences to obtain images with contrasts between soft tissue types. Different sequences can produce dramatically different appearance even for the same anatomy. For neuroimaging, a range of contrasts is commonly acquired to provide complementary information, such as T1-weighted contrast (T1w) for inspecting anatomy or T2-weighted contrast (T2w) for detecting abnormal fluids [1]. Registration of such images is critical when combining information across acquisitions, for example to gauge the damage induced by a stroke or to plan a brain-tumor resection. While rigid registration can be sufficient for aligning within-subject images acquired with the same sequence [2], images acquired with different sequences can undergo differential distortion due to effects such as eddy currents and susceptibility artifacts, requiring deformable registration [3]. Deformable registration is also important for morphometric analyses [4]–[6], which hinge on aligning images with an existing standardized atlas that typically has a different contrast [7]–[9]. Given the central importance of registration tasks within and across contrasts, and within and across subjects, the goal of this work is a learning-based framework for registration *agnostic* to MRI contrast: we propose a strategy for training networks that excel both within contrasts (e.g. between two T1w scans) *as well as* across contrasts (e.g. T1w to T2w), even if the test contrasts are not observed during training.

Classical registration approaches estimate a deformation field between two images by optimizing an objective that balances image similarity with field regularity [10]–[16]. While these methods provide a strong theoretical background and can yield good results, the optimization needs to be repeated for every new image pair, and the objective and optimization strategy typically need to be adapted to the image type. In contrast, learning-based registration uses datasets of images to learn a function that maps an image pair to a deformation field aligning the images [17]–[24]. These approaches achieve sub-second runtimes on a GPU and have the potential to improve accuracy and robustness to local minima. Unfortunately, they are limited to the MRI contrast available during training and therefore do not generally perform well on unobserved (new) image types. For example, a model trained on pairs of T1w and T2w images will not accurately register T1w to proton-density weighted (PDw) images. With a focus on neuroimaging, we remove this constraint of learning methods and design an approach that generalizes to *unseen* MRI contrasts at test time.

### A. Related Work

*1) Classical Methods:* Deformable registration has been widely studied [11], [12], [15], [16], [25]. Classical strategies implement an iterative procedure that estimates an optimal deformation field for each image pair. This involves maximizing an image-similarity metric, that compares the warped moving and fixed images, and a regularization term that encourages desirable deformation properties such as preservation of topology [10], [13]–[15]. Cost function and optimization strategies are typically chosen to suit a particular task. Simple metrics like mean squared error (MSE) or normalized cross-correlation (NCC) [12] are widely used and provide excellent accuracy for images of the same contrast [26].

For registration across MRI contrasts, metrics such as mutual information (MI) [27] and correlation ratio [28] are often employed, although the accuracy achieved with them is not on par with the within-contrast accuracy of NCC and MSE [29]. For some tasks, e.g. registering intra-operative ultrasound to MRI, estimating even approximate correspondences can be challenging [30], [31]. While they are not often used in neuroimaging, metrics based on patch similarity [32]–[36] and normalized gradient fields [37]–[39] outperform simpler metrics, e.g. on abdominal computer-tomography (CT). Other methods convert images to a supervoxel representation, which is then spatially matched instead of the images [40], [41]. Our work also employs geometric shapes, but instead of generating supervoxels from input images, we synthesize arbitrary patterns (and images) from scratch during training to encourage learning contrast-invariant features for spatial correspondence.

*2) Learning Approaches:* Learning-based techniques mostly use convolutional neural networks (CNNs) to learn a function that directly outputs a deformation field given an image pair. After training, evaluating this function is efficient, enabling fast registration. Supervised models learn to reproduce simulated warps or deformation fields estimated by classical methods [21], [22], [24], [42]–[44]. In contrast, unsupervised models minimize a loss similar to classical cost functions [17], [45]–[47] such as normalized MI (NMI) [48] for cross-contrast registration. In another cross-contrast registration paradigm, networks synthesize one contrast from the other, so that within-contrast losses can be used for subsequent nonlinear registration [29], [49]–[53]. These methods all depend on having training data of the target contrast. If no such data are available during training, models generally predict inaccurate warps at test time: a model trained on T1w-T1w pairs would fail when applied within unseen contrasts (e.g. T2w-T2w) or across unseen contrast combinations (e.g. T1w-T2w).

Recent approaches also use losses driven by label maps or sparse annotations (e.g. fiducials) for registering different imaging modalities labeled during training, such as T2w MRI and 3D ultrasound within the same subject [54], [55], or aiding existing formulations with auxiliary segmentation data [17], [56]–[58]. While these label-driven methods can boost registration accuracy compared to approaches using intensity-based loss functions, they are dependent on the limited annotated images available during training. Consequently, these approaches do not perform well on unobserved MRI contrasts.

Data-augmentation strategies expose a model to a wider range of variability than the training data encompasses, for example by randomly altering voxel intensities or applying deformations [59]–[62]. However, even these methods still need to sample training data acquired with the target contrast. Similarly, transfer learning can be used to extend a trained network to new contrasts but does not remove the need for training data with the target contrast [63]. Given the continuing development of new and improved MRI contrast types at ever higher field strengths, the reduction in accuracy evidenced by existing methods in the presence of novel image contrast becomes a limiting factor.

### B. Contribution

In this work we present `SynthMorph`, a general strategy for learning contrast-agnostic registration (Fig. 1). At test time, it can accurately register a wide variety of acquired images with MRI contrasts *unseen* during training. `SynthMorph` enables registration of real images both within and across contrasts, learning only from synthetic data that far exceed the realistic range of medical images. During training we synthesize images from label maps, whereas registration requires no label maps at test time. First, we introduce a generative model for random label maps of variable geometric shapes. Second, conditioned on these maps, or optionally given other maps of interest, we build on recent methods to synthesize images with arbitrary contrasts, deformations, and artifacts [64]. Third, the strategy enables us to use a contrast-agnostic loss that measures label overlap, instead of an image-based loss. This leads to two `SynthMorph` network variants (sm) that yield substantial generalizability, both capable of registering any contrast combination tested without retraining: `sm-shapes` trains without acquired data of *any* kind, matches classical state-of-the-art registration of neuroanatomical MRI, and outperforms learning baselines at cross-contrast registration. Variant `sm-brains` trains on images synthesized from brain segmentations only and substantially outperforms all classical and learning-based baselines tested.

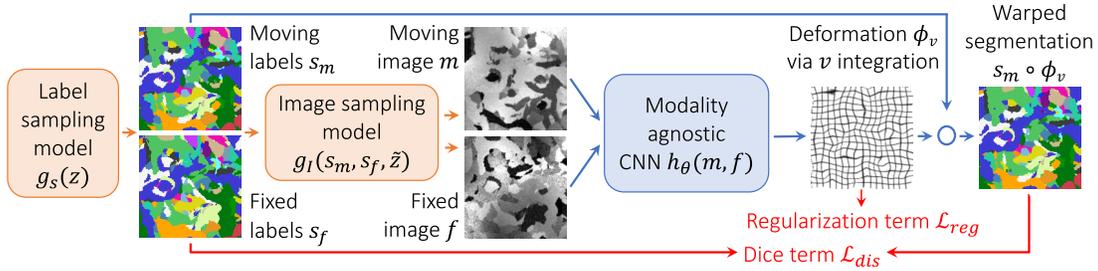

Fig. 1. Unsupervised learning strategy for contrast-agnostic registration. At every mini batch, we synthesize a pair of 3D label maps $\{s_m, s_f\}$ and the corresponding 3D images $\{m, f\}$ from noise distributions. The label maps are incorporated into a loss that is independent of image contrast.

This work builds on and extends a preliminary conference paper [65] presented at the IEEE International Symposium on Biomedical Imaging (ISBI) 2021. The extension includes a series of new experiments, new analyses of the framework and regularization, and a substantially expanded discussion. We also show that networks trained within the `SynthMorph` strategy generalize to new image types with MRI contrasts *unseen* at training. Our contribution focuses on neuroimaging but provides a general learning framework that can be used to *train* models across imaging applications and machine-learning techniques. Our code is freely available as part of the VoxelMorph library [66] and at https://w3id.org/synthmorph.

## II. METHOD

### A. Background

Let $m$ and $f$ be a moving and a fixed 3D image, respectively. We build on unsupervised learning-based registration frameworks and focus on deformable (non-linear) registration. These frameworks use a CNN $h_\theta$ with parameters $\theta$ that outputs the deformation $\phi_\theta = h_\theta(m, f)$ for image pair $\{m, f\}$.

At each training iteration, the network $h_\theta$ is given a pair of images $\{m, f\}$, and parameters are updated by optimizing a loss function $\mathcal{L}(\theta; m, f, \phi_\theta)$ similar to classical cost functions, using stochastic gradient descent. Typically, the loss contains an image dissimilarity term $\mathcal{L}_{dis}(m \circ \phi_\theta, f)$ that penalizes differences in appearance between the warped image and the fixed image, and a regularization term $\mathcal{L}_{reg}(\phi)$ that encourages smooth deformations:

$$\mathcal{L}(\theta; m, f, \phi_\theta) = \mathcal{L}_{dis}(m \circ \phi_\theta, f) + \lambda \mathcal{L}_{reg}(\phi_\theta), \quad (1)$$

where $\phi_\theta = h_\theta(m, f)$ is the network output, and $\lambda$ controls the weighting of the terms. Unfortunately, networks trained this way only predict reasonable deformations for images with contrasts and shapes similar to the data observed during training. Our framework alleviates this dependency.

### B. Proposed Method Overview

We strive for contrast invariance and robustness to anatomical variability by requiring no acquired training data, but instead synthesizing arbitrary contrasts and shapes from scratch (Fig. 1). We generate two paired 3D label maps $\{s_m, s_f\}$ using a function $g_s(z) = \{s_m, s_f\}$ described below, given random seed $z$. However, if anatomical labels are available, we can use these instead of synthesizing segmentation maps. We then define another function $g_I(s_m, s_f, \tilde{z}) = \{m, f\}$ (described below) that synthesizes two intensity volumes $\{m, f\}$ based on the maps $\{s_m, s_f\}$ and seed $\tilde{z}$.

This generative process resolves the limitations of existing methods as follows. First, training a registration network $h_\theta(m, f)$ using the generated images exposes it to arbitrary contrasts and shapes at each iteration, removing the dependency on a specific MRI contrast. Second, because we first synthesize label maps, we can use a similarity loss that measures label overlap independent of image contrast, thereby obviating the need for a cost function that depends on the contrasts being registered at that iteration. In our experiments, we use the (soft) Dice metric [67]

$$\mathcal{L}'_{dis}(\phi, s_m, s_f) = -\frac{2}{J} \sum_{j=1}^{J} \frac{|(s_m^j \circ \phi) \odot s_f^j|}{|(s_m^j \circ \phi) \oplus s_f^j|}, \quad (2)$$

where $s^j$ represents the one-hot encoded label $j \in \{1, 2, \ldots, J\}$ of label map $s$, and $\odot$ and $\oplus$ denote voxel-wise multiplication and addition, respectively.

While the framework can be used with any parameterization of the deformation field $\phi$, in this work we use a stationary velocity field (SVF) $v$, which is integrated within the network to obtain a diffeomorphism [11], [45], [68], that is invertible by design. We regularize $\phi$ using $\mathcal{L}_{reg}(\phi) = \frac{1}{2} \|\nabla \mathbf{u}\|^2$, where $\mathbf{u}$ is the displacement of the deformation field $\phi = Id + \mathbf{u}$.

### C. Generative Model Details

*1) Label Maps:* To generate input label maps with $J$ labels of random geometric shapes, we first draw $J$ smoothly varying noise images $p_j$ ($j \in \{1, 2, \ldots, J\}$) by sampling voxels from a standard distribution at lower resolution $r_p$ and upsampling to full size (Fig. 2). Second, each image $p_j$ is warped with a random smooth deformation field $\phi_j$ (described below) to obtain images $\tilde{p}_j = p_j \circ \phi_j$. Third, we create an input label map $s$ by assigning, for each voxel $k$ of $s$, the label $j$ corresponding to image $\tilde{p}_j$ that has the highest intensity, i.e. $s_k = \arg\max_j([\tilde{p}_j]_k)$.

Given a selected label map $s$, we generate two new label maps. First, we deform $s$ with a random smooth diffeomorphic

TABLE I

HYPERPARAMETERS. SPATIAL MEASURES ARE IN VOXELS. OUR IMAGES AND LABEL MAPS ARE $160 \times 160 \times 192$ VOLUMES. FOR FIELDS SAMPLED AT RESOLUTION $r$, WE OBTAIN THE VOLUME SIZE BY MULTIPLYING EACH DIMENSION BY $r$ AND ROUNDING UP. FOR EXAMPLE, A RESOLUTION OF $r = 1{:}40$ RELATIVE TO THE VOLUME SIZE $160 \times 160 \times 192$ WOULD BE EQUIVALENT TO WORKING WITH VOLUMES OF SIZE $4 \times 4 \times 5$ transformation $\phi_m$ (described below) using nearest-neighbor interpolation to produce the moving segmentation map $s_m = s \circ \phi_m$. An analogous process yields the fixed map $s_f$.

| Hyperparameter | $\lambda$ | $r_p$ | $b_p$ | $a_\mu$ | $b_\mu$ | $a_\sigma$ | $b_\sigma$ | $r_B$ | $b_B$ | $b_K$ | $\sigma_\gamma$ | $r_v$ | $b_v$ |
|---|---|---|---|---|---|---|---|---|---|---|---|---|---|
| Value | 1 | 1:32 | 100 | 25 | 225 | 5 | 25 | 1:40 | 0.3 | 1 | 0.25 | 1:16 | 3 |

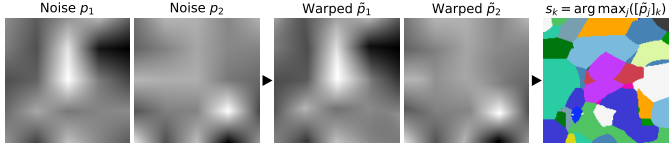

**Fig. 2.** Generation of input label maps. Smooth 3D noise images $p_j$ ($j \in \{1, 2, \ldots, J\}$) are sampled from a standard distribution, then warped by random deformations $\phi_j$ to cover a range of scales and shapes. We synthesize a label map $s$ from the warped images $\tilde{p}_j = p_j \circ \phi_j$: for each voxel $k$ of $s$, we assign label $j$ corresponding to image $\tilde{p}_j$ where $k$ has the highest intensity $j$, i.e. $s_k = \arg\max_j([\tilde{p}_j]_k)$. We use $J = 26$.

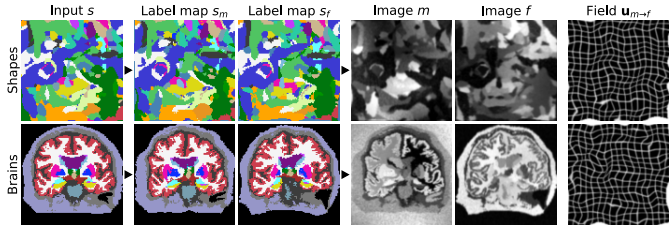

**Fig. 3.** Data synthesis. Top: from random shapes. Bottom: if available, from anatomical labels. We generate a pair of label maps $\{s_m, s_f\}$ and from them images $\{m, f\}$ with arbitrary contrast. The registration network then predicts the displacement $\mathbf{u}_{m \to f}$. If anatomical labels are used, we generate $\{s_m, s_f\}$ from separate subjects.

Alternatively, if segmentations are available for the anatomy of interest, such as the brain, we randomly select and deform input label maps instead of synthesizing them (Fig. 3). To generate two different images, we could start by using a single segmentation twice, or two separate ones. In this work, we sample separate brain label maps as this captures more realistic variability in the correspondences that the registration network has to find. In contrast, for sm-shapes training, we use a single label map $s$ as input twice to ensure that topologically consistent correspondences exist.

*2) Synthetic Images:* From the pair of label maps $\{s_m, s_f\}$, we synthesize gray-scale images $\{m, f\}$ building on generative models of MR images used for Bayesian segmentation [69]–[72] (Fig. 3). We extend a publicly available model [64] to make it suitable for registration, which, in contrast to segmentation, involves the efficient generation of pairs of images (Section II-D.3). Given a segmentation map $s$, we draw the intensities of all image voxels that are associated with label $j$ as independent samples from the normal distribution $\mathcal{N}(\mu_j, \sigma_j^2)$. We sample the mean $\mu_j$ and standard deviation (SD) $\sigma_j$ for each label from continuous distributions $\mathcal{U}(a_\mu, b_\mu)$ and $\mathcal{U}(a_\sigma, b_\sigma)$, respectively, where $a_\mu$, $b_\mu$, $a_\sigma$, and $b_\sigma$ are hyperparameters. To simulate partial volume effects [73], we convolve the image using an anisotropic Gaussian kernel $K(\sigma_{i=1,2,3})$ where $\sigma_{i=1,2,3} \sim \mathcal{U}(0, b_K)$.

We further corrupt the image with a spatially varying intensity-bias field $B$ [74], [75]. We independently sample the voxels of $B$ from a normal distribution $\mathcal{N}(0, \sigma_B^2)$ at lower resolution $r_B$ relative to the full image size (described below), where $\sigma_B \sim \mathcal{U}(0, b_B)$. We upsample $B$ to full size, and take the exponential of each voxel to yield non-negative values before we apply $B$ using element-wise multiplication. We obtain the final images $m$ and $f$ through min-max normalization and contrast augmentation through global exponentiation, using a single normally distributed parameter $\gamma \sim \mathcal{N}(0, \sigma_\gamma^2)$ for the entire image such that $m = \tilde{m}^{\exp(\gamma)}$, where $\tilde{m}$ is the normalized moving image, and similarly for the fixed image (Fig. 3).

*3) Random Transforms:* We obtain the transforms $\phi_j$ ($j = 1, 2, \ldots, J$) for noise image $p_j$ by integrating random SVFs $v_j$ [11], [45], [46], [68]. We draw each voxel of $v_j$ as an independent sample of a normal distribution $\mathcal{N}(0, \sigma_j^2)$ at lower resolution $r_p$, where $\sigma_j \sim \mathcal{U}(0, b_p)$ is sampled uniformly, and each SVF is integrated and upsampled to full size. Similarly, we obtain the transforms $\phi_m$ and $\phi_f$ based on hyperparameters $r_v$ and $b_v$ for sm-brains. For sm-shapes, we sample several SVFs $v_m \sim \mathcal{N}(0, \sigma_v^2)$ at resolutions $r_v \in \{1{:}8, 1{:}16, 1{:}32\}$, drawing a different $\sigma_v$ for each to synthesize a more complex deformation, since the fixed and moving images are based on the same input label map. The upsampled SVFs are then combined additively, and a similar procedure yields $v_f$.

### D. Implementation Details

*1) Hyperparameters:* The generative process requires a number of parameters. During training, we sample these based on the hyperparameters presented in Table I. Their values are *not* chosen to mimic realistic anatomy or a particular MRI contrast. Instead, we select hyperparameters visually to yield shapes and contrasts that far exceed the range of realistic medical images, to force our networks to learn generalizable features that are independent of the characteristics of a specific contrast [59]. We thoroughly analyze the impact of varying hyperparameters in our experiments.

*2) Architecture:* The models implement the network architecture used in the VoxelMorph library [17], [45]: a convolutional U-Net [60] predicts an SVF $v_\theta$ from the input $\{m, f\}$. As shown in Fig. 4, the encoder has 4 blocks consisting of a stride-2 convolution and a LeakyReLU layer (parameter 0.2), that each halve the resolution relative to the inputs.

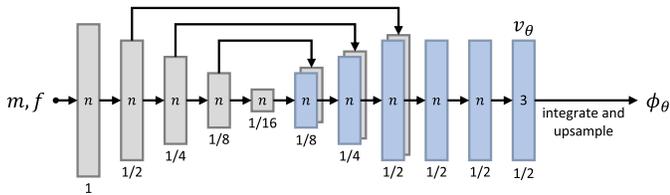

**Fig. 4.** U-Net architecture of $\phi_\theta = h_\theta(m,f)$. Each block of the encoder features a 3D convolution with $n = 256$ filters and a LeakyReLU layer (0.2). Stride-2 convolutions each halve the resolution relative to the input. In the decoder, each convolution is followed by an upsampling layer and a skip connection (long arrows). The SVF $v_\theta$ is obtained at half resolution, yielding the warp $\phi_\theta$ after integration and upsampling. All kernels are of size $3 \times 3 \times 3$. The final layer uses $n = 3$ filters.

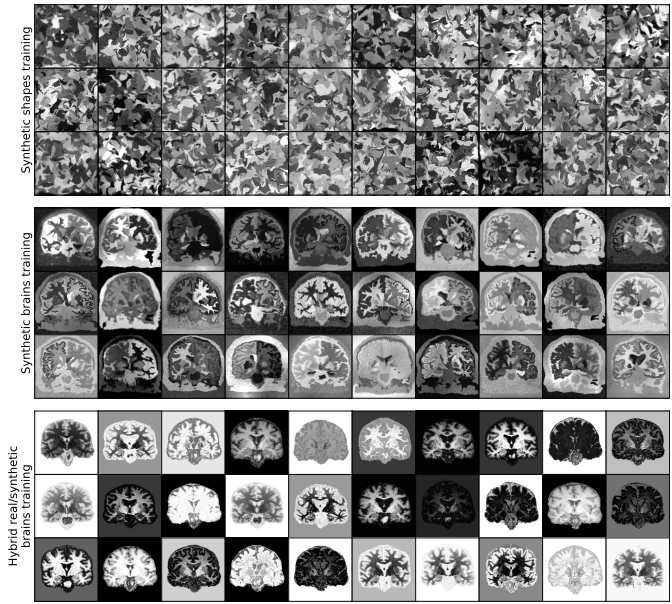

**Fig. 5.** Synthetic training data. Top: random geometric shapes synthesized from noise distributions. Center: arbitrary contrasts synthesized from brain segmentations. Bottom: hybrid synthesis requiring acquired MRI for contrast augmentation using smooth random lookup tables.

The decoder features 3 blocks that each include a stride-1 convolution, an upsampling layer, and a skip connection to the corresponding encoder block. We obtain the SVF $v_\theta$ after 3 further convolutions at half resolution, and the warp $\phi_\theta$ after integration and upsampling.

All convolutions use $3 \times 3 \times 3$ kernels. We use a default network width of $n = 256$ unless stated otherwise. While the last layer of all networks employs $n = 3$ filters, we reduce the width to $n = 64$ for the parameter sweeps of Section III-G and the analysis of feature maps in Fig. 10 and Fig. 12, to lower the computational burden and memory requirements and thereby enable us to perform the analyses within our computational resources. We expect the results to be generally applicable as we use the same synthesis and registration architecture, while higher network capacities typically improve accuracy as long as the training set is large enough.

*3) Implementation:* We implement our networks using TensorFlow/Keras [76]. We integrate SVFs using a GPU version [45], [46] of the *scaling and squaring* algorithm with 5 steps [11], [68]. Training uses the Adam optimizer [77] with a batch size of one registration pair and an initial learning rate of $10^{-4}$, that we decrease to $10^{-5}$ in case of divergence. We train each model until the Dice metric converges in the synthetic training set, typically for $4 \times 10^5$ iterations.

To generate pairs of images with high variability for registration, we extend a model [64] implemented for a single-input segmentation task. First, we improve efficiency to meet the increased computational demand. For example, we replace smoothing operations based on 3D convolutions by 1D convolutions with separated Gaussian kernels. We also integrate spatial augmentation procedures such as random axis flipping into a single deformation field, enabling their application as part of one interpolation step. We also implement an interpolation routine with fill-value-based extrapolation on the GPU. The fill value enables extrapolating with zeros instead of repeating voxels where the anatomy extends to the edge of the image, making the spatial augmentation more realistic.

Second, we add to the data augmentation within the model by expanding random axis flipping to all three dimensions, and by drawing a separate smoothing kernel for each dimension of space enabling randomized anisotropic blurring. We implement a more complex warp synthesis that generates and combines SVFs at multiple spatial resolutions. We also extend most augmentation steps to vary across batches, thereby increasing variability.

Third, we simplify the code to improve its maintainability and reusability. We use the external VoxelMorph and Neurite libraries to avoid code duplication. We update the model to support the latest TensorFlow version to benefit from the full set of features including batch profiling and debugging in eager execution mode.

## III. Experiments

We evaluate network variants trained with the proposed strategy and compare their performance to several baselines. The test sets include a variety of image contrasts and levels of processing to assess method robustness. Our goal is for `SynthMorph` to achieve unprecedented generalizability to new contrasts among neural networks, matching or exceeding the accuracy of all classical and learning methods tested.

### A. Data

While `SynthMorph` training involves data synthesized from label maps that vary widely beyond realistic ranges, all tests and method comparisons use only acquired MRI data.

*1) Datasets:*

*a) OASIS, HCP-A, BIRN:* We compile 3D brain-MRI datasets from the Open Access Series of Imaging Studies (OASIS) [78] and the Human Connectome Project Aging Project (HCP-A) [79], [80]. OASIS includes T1w MPRAGE acquired at 1.5 T with ∼(1 mm)³ resolution. HCP-A includes both T1w MEMPRAGE [81] and T2w T2SPACE [82] scans acquired at 3 T with 0.8 mm isotropic resolution. We also use PDw 1.5-T BIRN [83] scans from 8 subjects, which include manual brain segmentations. Fig. 6 shows typical image examples.

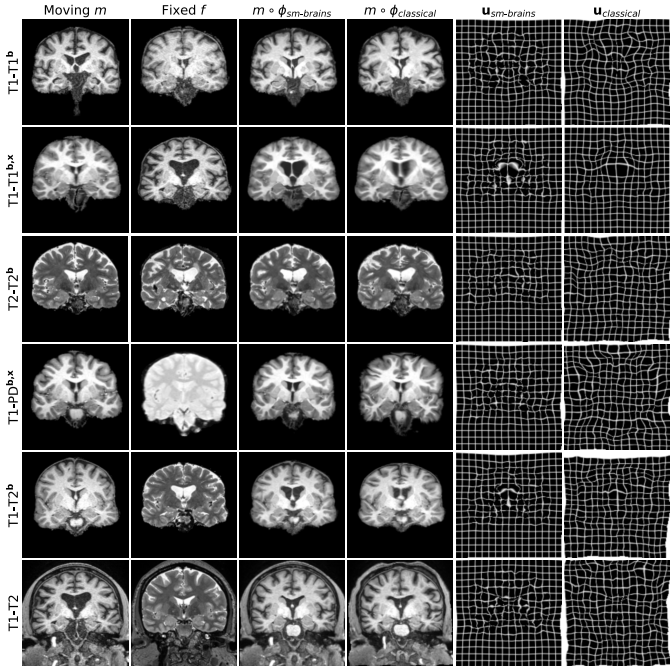

Fig. 6. Typical results for `sm-brains` and classical methods. Each row shows an image pair from the datasets indicated on the left. The letters **b** and **x** mark skull-stripping and registration across datasets (e.g. OASIS and HCP-A), respectively. We show the best classical baseline: `NiftyReg` on the 1st, `ANTs` on the 2nd, and `deedsBCV` on all other rows.

*b) UKBB, GSP:* We obtain 7000 skull-stripped T1w scans acquired at 3 T field strength. Of these, we source 5000 MPRAGE images with 1 mm isotropic resolution from the UK Biobank (UKBB) [84] and 2000 MEMPRAGE [81] scans with 1.2 mm isotropic resolution from the Brain Genomics Superstruct Project (GSP) [85].

*c) Multi-FA, Multi-TI:* We compile a series of spoiled gradient-echo (FLASH) [86] images for flip angles (FA) varied between 2° and 40° in 2° steps. For each of 20 subjects, we obtain contrasts ranging from PDw to T1w using the steady-state signal equation with acquired parametric maps (T1, T2*, PD) and sequence parameters: repetition time (TR) 20 ms, echo time (TE) 2 ms. Equivalently, we compile a series of MPRAGE images for inversion times (TI) varied between 300 ms and 1000 ms in steps of 20 ms. For each of 20 subjects, we fit MPRAGE contrasts based on MP2RAGE [87] echoes acquired with parameters: TR/TE 5000/2.98 ms, $TI_1/TI_2$ 700/2500 ms, FA 4°. Fig. 9 shows typical examples of these data.

*d) Buckner40:* We derive 40 distinct-subject segmentations with brain and non-brain labels from T1w MPRAGE scans of the Buckner40 dataset [88], a subset of the fMRIDC structural data [89].

*e) Cardiac MRI:* We gather cine-cardiac MRI datasets from 33 subjects [90]. Each frame is a stack of thick 6-13 mm slices with ∼(1.5 mm)² in-plane resolution. The data include manually drawn contours outlining the endocardial and epicardial walls of the left ventricle. Fig. 15 shows representative frames.

*2) Processing:* As we focus on deformable registration, we map all brain images into a common $160 \times 160 \times 192$ affine

|  | Moving | Fixed | Subject pairs | Experiment |
|---|---|---|---|---|
| T1-T1**b** | OASIS | OASIS | 50 | 1 |
| T1-T1**b** | HCP-A | HCP-A | 50 | 3 |
| T1-T1**b,x** | OASIS | HCP-A | 50 | 1 |
| T2-T2**b** | HCP-A | HCP-A | 50 | 1 |
| T1-PD**b,x** | OASIS | BIRN | 8 | 1 |
| T1-T2**b** | HCP-A | HCP-A | 50 | 1 |
| T1-T2 | HCP-A | HCP-A | 50 | 1 |

space [4], [91] at 1 mm isotropic resolution. Unless manual segmentations are available, we derive brain and non-brain labels for skull-stripping and evaluation using the contrast-adaptive `SAMSEG` [6] method.

For each subject of the multi-FA and multi-TI datasets, we derive brain labels from a single acquired T1w image using FreeSurfer [4], ensuring identical labels across all MRI contrasts obtained for the subject.

We resample all cardiac frames to $256 \times 256 \times 112$ volumes with isotropic-1-mm voxels and transfer the manual contours into the same space.

*3) Dataset Use:*

*a) Training:* We use the Buckner40 label maps for data synthesis (Fig. 5) and `SynthMorph` training. For the learning baselines, we use T1w and T2w images from 100 HCP-A subjects, and all T1w images from GSP and UKBB.

*b) Validation:* For hyperparameter tuning and monitoring model training, we use 10 registration pairs for each of the OASIS, HCP-A and BIRN contrast pairings described below. These subjects do not overlap with the training set.

*c) Test:* Table II provides an overview of the contrast combinations compiled from OASIS, HCP-A, and BIRN. Except for the 8 PDw BIRN images, the subjects do not overlap with the training or validation sets. We also use the multi-FA, multi-TI and cardiac images for testing; none of these data are used in training or validation.

### B. Baselines

We test classical registration with `ANTs` (SyN) [12] using recommended parameters [92] for the NCC similarity metric within contrast and MI across contrasts. We test `NiftyReg` [13] with the default cost function (NMI) and recommended parameters, and we enable its diffeomorphic model with SVF integration as in our approach. Both `ANTs` and `NiftyReg` are optimized for neuroimaging applications, leading to appropriate parameters for our tasks. We also run the `deedsBCV` [93] patch-similarity method, which we tune for neuroimaging. To match the spatial scales of brain structures, we reduce the default grid spacing, search radius and quantization step to $6 \times 5 \times 4 \times 3 \times 2$, $6 \times 5 \times 4 \times 3 \times 2$, and $5 \times 4 \times 3 \times 2 \times 1$, respectively, improving registration in our experiments.

As a learning baseline, we train `VoxelMorph` (vm), using an image-based NCC loss and the same architecture as SynthMorph, on 100 skull-stripped T1w images from HCP-A that do not overlap with the validation set. Similarly, we train another model with NMI as a loss on random combinations of 100 T1w and 100 T2w images. This exposes each model to 9900 different cross-subject registration pairs, and `vm-nmi` to T1w-T1w, T1w-T2w and T2w-T2w contrast pairings (both contrasts were acquired from the same 100 subjects). Following the original `VoxelMorph` implementation [17], we train these baseline networks without data augmentation, with the exception of randomized axis flipping.

While we compare to learning baselines following their original implementation [17], we also investigate if the performance of these methods can be further improved. First, we retrain the baseline model adding a further 7000 T1w images from UKBB and GSP to the training set to evaluate whether the original finding that 100 images are sufficient [17] holds true in our implementation, or whether the greater anatomical variability would promote generalizability across contrasts or datasets (`vm-ncc-7k`).

Second, we explore to what extent augmentation can improve accuracy, by retraining `vm-ncc` with 100 T1w images, while augmenting the input images with random deformations as for `sm-brains` training (`vm-ncc-aug`).

Third, we train a new `hybrid` method using extreme contrast augmentation to explore if more variability in the training contrasts would help the network generalize (Fig. 5). At every iteration, we sample a registration pair from 100 T1w images and pass it to the similarity loss, while the network inputs each undergo an arbitrary gray-scale transformation: we uniformly sample a random lookup table (LUT) from $\mathcal{U}(0, 255)$, remapping the intensities $\{0, \ldots, 255\}$ to new values of the same set. We smooth this LUT using a Gaussian kernel $L(\sigma_L = 64)$.

Fourth, the synthesis enables supervised training *if* the moving and fixed label maps $\{s_m, s_f\}$ are generated from the same input label map, so that the net warp is known. We analyze whether knowledge of the synthetic net warp can improve accuracy, by training models with the same architecture using an MSE loss between the synthesized and predicted SVFs $v$ (sup-svf) or deformation fields $\phi$ (sup-def), respectively. As for `sm-shapes`, we draw the SVFs $\{v_m, v_f\}$ at several resolutions $r_v \in \{1{:}8, 1{:}16, 1{:}32\}$ to synthesize a more complex deformation since we use a single brain segmentation map as input to ensure that a topologically consistent spatial correspondence exists.

### C. SynthMorph Variants

For image-data and shape-agnostic training (`sm-shapes`), we sample $\{s_m, s_f\}$ by selecting one of 100 random-shape segmentations $s$ at each iteration, and synthesizing two separate image-label pairs from it. Each $s$ contains $J = 26$ labels that we include in the loss $\mathcal{L}_{dis}$. Since brain segmentations are often available, even if not for the target MRI contrast, we train another network on the Buckner40 anatomical labels instead of shapes (`sm-brains`). In this case, we sample $\{s_m, s_f\}$ from two distinct label maps at each iteration and further deform them using synthetic warps. We optimize the $J = 26$ largest brain labels in $\mathcal{L}_{dis}$, similar to what `VoxelMorph` does for validation [17] (see below).

### D. Validation Metrics

To measure registration accuracy, we propagate the moving labels using the predicted warps and compute the Dice metric $D$ [94] across a representative set of brain structures: amygdala, brainstem, caudate, ventral DC, cerebellar white matter and cortex, pallidum, cerebral white matter (WM) and cortex, hippocampus, lateral ventricle, putamen, thalamus, $3^{rd}$ and $4^{th}$ ventricle, and choroid-plexus. We average scores of bilateral structures. In addition to volumetric Dice overlap, we evaluate the mean symmetric surface distance $S$ (MSD) between contours of the same moved and fixed labels. We also compute the proportion of voxels where the warp $\phi$ folds, i.e. $\det(J_\phi) \le 0$ for voxel Jacobian $J_\phi$.

### E. Experiment 1: Baseline Comparison

*1) Setup:* For each contrast, we run experiments on 50 held-out image pairs, where each image is from a different subject, except for T1w-PDw pairs, of which we have eight. To assess robustness to non-brain structures, we evaluate registration within and across datasets, with and without skull-stripping, using datasets of the same size. We determine whether the mean differences between methods are significant using paired two-sided t-tests.

*2) Results:* Fig. 5 shows examples of `SynthMorph` training data, and Fig. 6 shows typical registration results. Fig. 7 compares registration accuracy across structures to all baselines, in terms of Dice overlap (Fig. 7a) and MSD (Fig. 7b). By exploiting the anatomical information in a set of brain labels, `sm-brains` achieves the best accuracy across all datasets, even though no real MR images are used during training. First, `sm-brains` outperforms classical methods on all tasks by at least 2.4 Dice points, and often much more ($p < 0.0003$ for T1w-PDw, $p < 4 \times 10^{-15}$ for all other tasks). Second, it exceeds the state-of-the-art accuracy of `vm-ncc` for T1w-T1w registration, which is trained on T1w images, by at least 0.6 Dice points ($p < 6 \times 10^{-6}$). Importantly, across contrasts `sm-brains` outperforms all other methods, demonstrating its ability to generalize to contrasts. Compared especially to baseline learning methods, which cannot generalize to contrasts *unseen* during training, `sm-brains` leads by up to 45.1 points ($p < 6 \times 10^{-7}$ for all cross-contrast tasks). Compared to classical methods, the proposed method outperforms by 2.9 or more points ($p < 0.0003$ for T1w-PDw, $p < 2 \times 10^{-17}$ for other cross-contrast tasks).

The shape and contrast-agnostic network `sm-shapes` matches the performance of the best classical method for each dataset except T1w-T1w registration, where it slightly underperforms ($p < 8 \times 10^{-11}$), despite never having been exposed to either imaging data or even neuroanatomy. Like `sm-brains`, `sm-shapes` generalizes well to multi-contrast registration, matching or exceeding the accuracy of all baselines, and by significant margins compared to learning baselines ($p < 8 \times 10^{-7}$ for T1w-PDw, $p < 2 \times 10^{-17}$ otherwise).

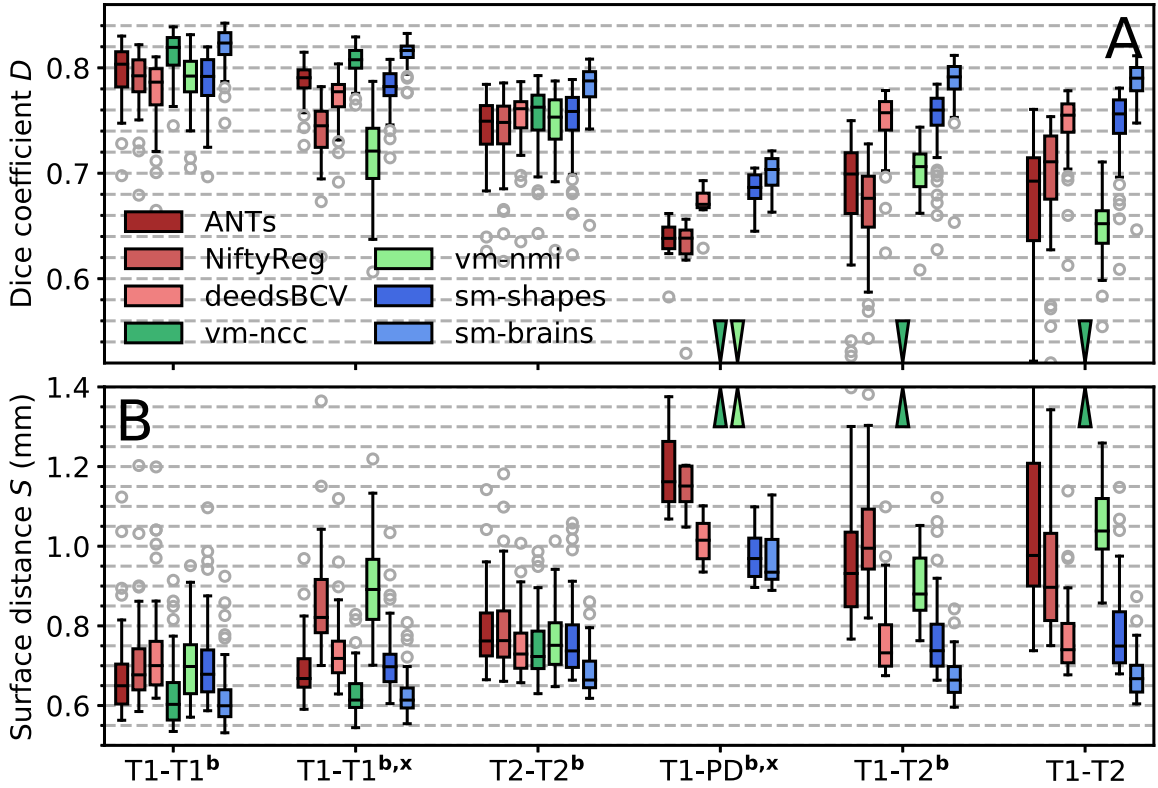

Fig. 7. Registration accuracy compared to baselines as (a) volume overlap $D$ using the Dice metric, and (b) mean symmetric surface distance $S$ between label contours. Each box shows mean accuracy over anatomical structures for 50 test-image pairs across distinct subjects (8 for PD). The letters **b** and **x** indicate skull-stripping and registration across datasets (e.g. OASIS-HCP), respectively. Arrows indicate values off the chart.

The baseline learning methods `vm-ncc` and `vm-nmi` perform well and clearly match or outperform classical methods at contrasts similar to those used in training. However, as expected, these approaches break down when tested on a pair of new contrasts that were not sampled during training, such as T1w-PDw. Similarly, `vm-ncc` and `vm-nmi` achieve slightly lower accuracy on image pairs that are not skull-stripped.

While MSD can be more sensitive than Dice overlap at structure boundaries, our analysis of surface distances yields a similar overall ranking between methods (Fig. 7b). Importantly, `sm-brains` achieves the lowest MSD for all contrasts, typically 0.7 mm or less, which is below the voxel size. Within contrasts, `sm-brains` outperforms classical methods by at least 0.06 mm ($p < 2 \times 10^{-9}$), surpassing all baselines tested across contrasts ($p < 0.04$ for T1w-PDw, $p < 10^{-10}$ for the other tasks).

Exposing the baseline models to a much larger space of deformations at training does not result in a statistically significant increase of accuracy for T1w-to-T1w registration within OASIS (Fig. 8a). For `vm-ncc-aug`, accuracy across T1w datasets (OASIS-HCP, $p < 0.007$) and T2w-to-T2w accuracy ($p < 0.03$) decrease by 0.1 Dice point relative to `vm-ncc`. For `vm-ncc-7k`, accuracy across T1w datasets increases by 0.1 point ($p < 0.04$), with no significant change for T2w-to-T2w registration, but overall these 0.13% changes are negligible. Similar to `vm-ncc`, these models do not generalize to unseen pairings across contrasts, under-performing `sm-brains` by 42.9 or more points (Fig. 8a, $p < 10^{-8}$).

Augmenting T1w image contrast using random LUTs (Fig. 5) substantially enhances performance across contrasts for `hybrid` compared to `vm-ncc` ($p < 2 \times 10^{-7}$), exceeding the supervised models by up to 6.1 Dice points ($p < 0.009$ for T1w-PDw, $p < 3 \times 10^{-15}$ for all other tasks). However, the increased contrast robustness comes at the expense of a drop of 0.5-1.9 Dice points within contrasts relative to `vm-ncc` ($p < 0.0002$), while `sm-brains` outperforms `hybrid` by at least 2.4 points within ($p < 6 \times 10^{-17}$) and 4.5 points across contrasts ($p < 10^{-5}$ for T1w-PDw, otherwise $p < 10^{-23}$). We also investigate lower kernel widths $\sigma_L < 64$, but find these to negatively impact accuracy and therefore do not include them in the graph: reducing $\sigma_L$ introduces noise in the image, indicating the importance of LUT smoothness.

Finally, the supervised networks `sup-def` and `sup-vel` achieve the lowest accuracy for within-contrast registration ($p < 0.02$) and consistently under-perform their unsupervised counterpart `sm-brains` by 6.8-10.7 points across all contrast combinations ($p < 0.0001$ for T1w-PDw, $p < 3 \times 10^{-26}$ for all other tasks). As for the main baseline comparison, measurements of the mean surface distance in Fig. 8b result

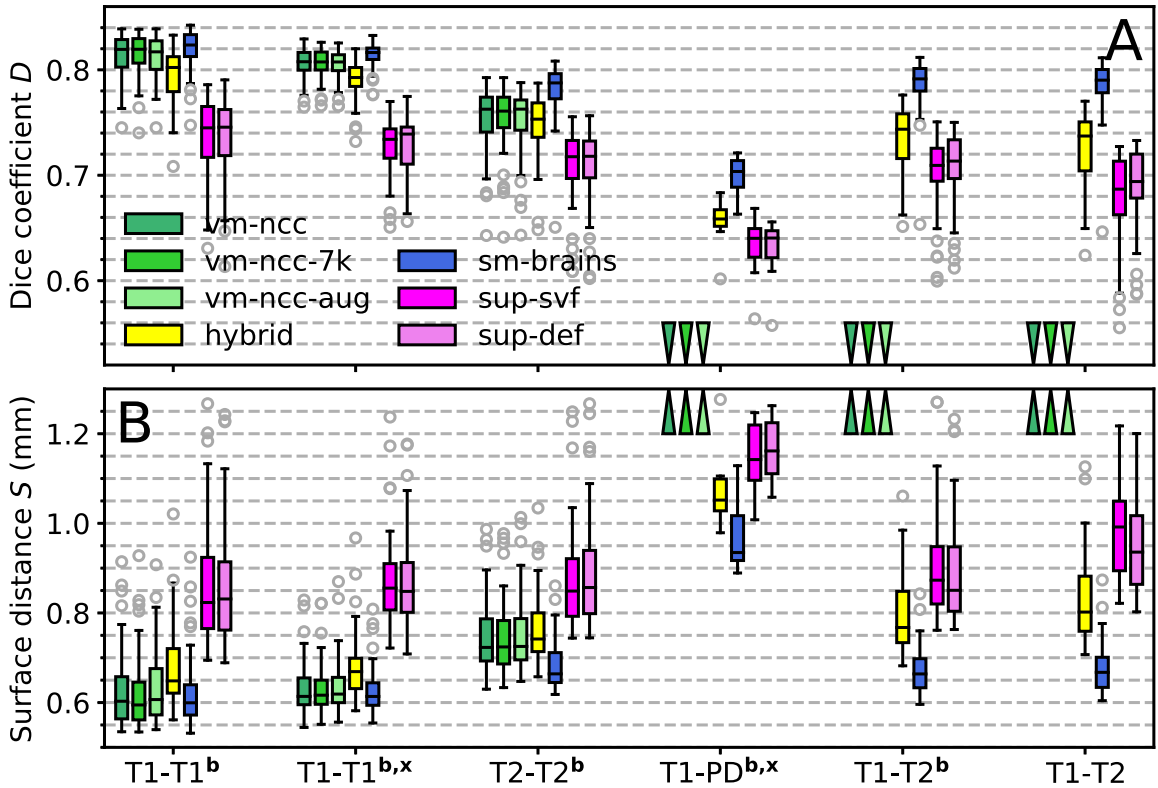

Fig. 8. Registration accuracy of method variations as (a) volume overlap *D* using the Dice metric, and (b) mean symmetric surface distance *S* between label contours. Each box shows mean accuracy over anatomical structures for 50 test-image pairs across distinct subjects (8 for PD). The letters **b** and **x** indicate skull-stripping and registration across datasets (e.g. OASIS-HCP), respectively. Arrows indicate values off the chart.

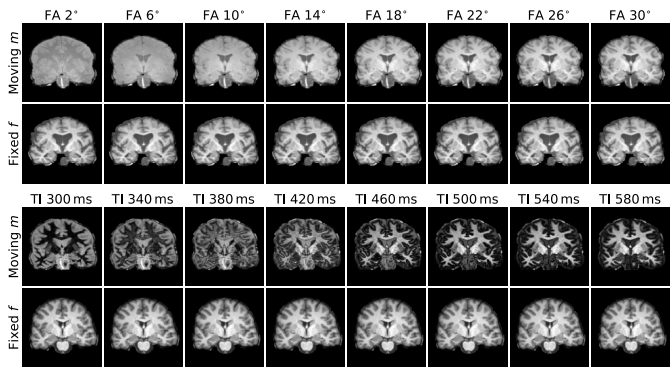

Fig. 9. Real MRI-contrast pairs used to assess network invariance. Top: we obtain FLASH images progressing from PDw (top left) to T1w for the same brain by varying FA using the steady-state signal equation with acquired parametric maps (T1, T2*, PD). Bottom: we obtain MPRAGE contrasts with varying TI by fitting intensities based on a dual-echo MP2RAGE scan (T1$_1$/T1$_2$ 700/2500 ms). For each of 10 subject pairs, we register a range of moving contrasts to a fixed T1w image.

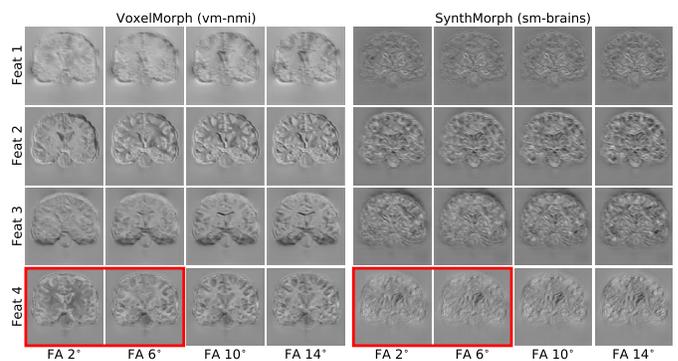

Fig. 10. Representative features of the last network layer before the stationary velocity field is formed, in response to evolving MRI contrasts from the same subject. Left: `VoxelMorph` using normalized mutual information (NMI) exhibits high variability of the same feature response across different input contrasts for the same brain, e.g. in the red box. Right: contrast-invariant `SynthMorph` (`sm-brains`). For this analysis, both networks use the same architecture with $n = 64$ filters per layer.

in a similar ranking between method variations, at comparable significance levels.

In our experiments, learning-based models require less than 1 second per 3D registration on an Nvidia Tesla V100 GPU. Using the recommended settings, `NiftyReg` and `ANTs` typically take ~0.5 h and ~1.2 h on a 3.3-GHz Intel Xeon CPU, respectively, whereas `deedsBCV` requires ~3 min.

### F. Experiment 2: Contrast Invariance

In this experiment we evaluate registration accuracy as a function of gradually varying MRI contrast and measure robustness to new image types by analyzing the variability of network features across these contrasts.

*1) Setup:* To assess network feature invariance to MRI contrast, we perform the following procedure for 10 pairs

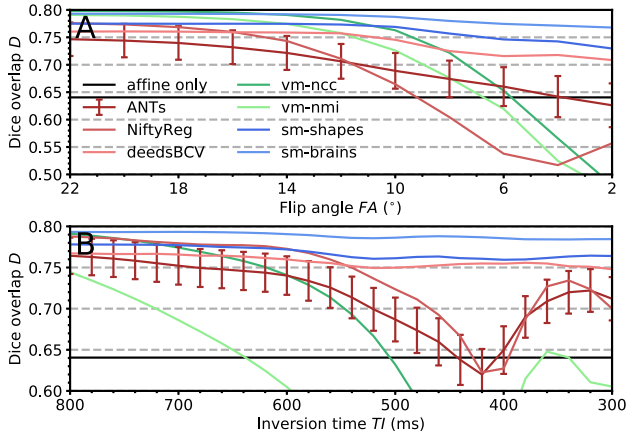

Fig. 11. Accuracy as a function of moving-image contrast across 10 realistic (a) FLASH and (b) MPRAGE image pairs. In each registration, the fixed image has the same T1w contrast. The moving image becomes decreasingly T1w towards the right. Being comparable across methods, error bars are shown for `ANTs` only and indicate the standard error of the mean over subjects.

of separate subjects, where each subject is only considered once and, thus, registered to a different fixed image. Given each such pair, we run a separate registration between each of the multi-FA contrasts for the moving subject and the most T1w-like contrast (FA 40°) of the fixed subject. For each pair of subjects, we measure accuracy with all tested methods as well as the variability of the features of the last network layer, before the SVF is formed, across input pairs. Specifically, we compute the root-mean-square difference $d$ (RMSD) between the layer outputs of the first and all other contrast pairs over space, averaged over contrasts, features, and subjects. For efficiency, we restrict the moving images for this analysis to the subsets of FAs and TIs that undergo the largest changes in contrast, i.e. FAs from 2 to 30° (4° steps) and TIs from 300 to 600 ms (40-ms steps).

*2) Results:* Fig. 11 compares registration accuracy as a function of the moving-image MRI contrast for baseline methods and `SynthMorph`. In both the multi-FA and the multi-TI data, we obtain broadly comparable results for all methods when the moving and fixed image have T1w-like contrast. However, the performance of `ANTs`, `NiftyReg` and learning baselines decreases with increasing contrast differences, whereas `SynthMorph` remains largely unaffected.

Fig. 12 shows the variability of the response of each network layer to varying MRI contrast of the same anatomy (shown in Fig. 9). Compared to `VoxelMorph`, the feature variability within the deeper layers is significantly lower for the `SynthMorph` models. Fig. 10 illustrates this result, containing example feature maps extracted from the last network layer before the SVF is formed.

Overall, `SynthMorph` models exhibit substantially less variability in response to contrast changes than all other methods tested, indicating that the proposed strategy does indeed encourage contrast invariance.

### G. Experiment 3: Hyperparameter Analyses

*1) Setup:* We explore the effect of various hyperparameters on registration performance using 50 skull-stripped HCP-A

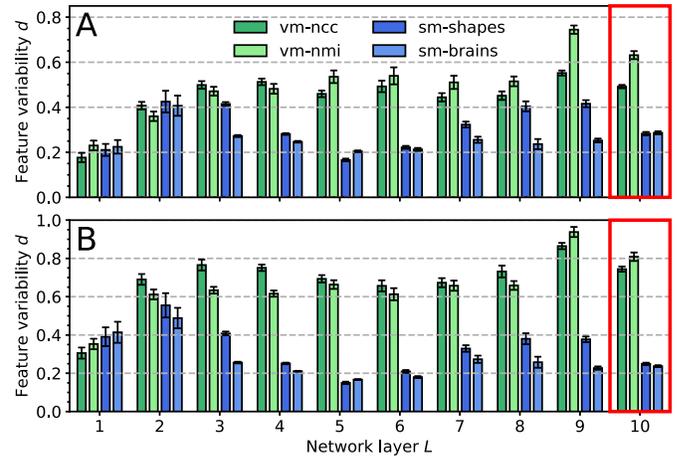

Fig. 12. Feature variability for registration across (a) FLASH and (b) MPRAGE contrasts from 10 distinct subject pairs. We use normalized RMSD $d$ between each contrast and the most T1w-like, averaged over contrasts, features, subjects. All models use the same architecture with $n = 64$ filters per layer. `SynthMorph` variants exhibit the least variability in the deeper layers (red boxes). Error bars show the standard error of the mean over features.

T1w pairs that do not overlap with the test set. First, we train with regularization weights $\lambda \in [0, 10]$ and evaluate accuracy across: (1) all brain labels and (2) only the largest 26 (bilateral) structures optimized in $\mathcal{L}_{dis}$. Second, we train variants of our model with varied deformation range $b_v$, image smoothness $b_K$, number of features $n$ per layer (network width), bias-field range $b_B$, gamma-augmentation strength $\sigma_\gamma$ and relative resolutions $r$. Third, for the case that brain segmentations are available (`sm-brains`), we analyze the effect of training with full-head labels, brain labels only, or a mixture of both. Unless indicated, we test all hyperparameters using $n = 64$ convolutional filters per layer. For comparability, both `SynthMorph` variants use SVFs $\{v_m, v_f\}$ sampled at a single resolution $r_v$.

*2) Results:* Fig. 13 shows registration performance for various training settings. Variant `sm-brains` performs best at low deformation strength $b_v$, when label maps $s$ from two different subjects are used at each iteration (Fig. 13a), likely because the differences between distinct subjects already provide significant variation. For $\{s_m, s_f\}$, a larger value of $b_v = 3$ is optimal due to the lacking inter-subject deformation, since we generate $\{s_m, s_f\}$ from a single segmentation $s$.

Random blurring of the images $\{m, f\}$ improves robustness to data with different smoothing levels, with optimal accuracy at $b_K \approx 1$ (Fig. 13b). Higher numbers of filters $n$ per convolutional layer boost the accuracy at the cost of increasing training times (Fig. 13c), indicating that richer networks better capture and generalize from synthesized data. We identify the optimum bias-field cap and gamma-augmentation SD as $b_B = 0.3$ and $\sigma_\gamma = 0.25$, respectively. We obtain the highest accuracy when we sample the SVF and bias field at relative resolutions $r_v = 1{:}16$ and $r_B = 1{:}40$, respectively (Fig. 13f). Finally, training on full-head as compared to skull-stripped images has little impact on accuracy (not shown).

Fig. 14a shows that with decreasing regularization, accuracy increases for the large structures used in $\mathcal{L}_{dis}$. When we EFFECT OF 3D REGISTRATION ON MEAN SYMMETRIC SURFACE DISTANCE (MSD) BETWEEN MANUALLY DRAWN CONTOURS OF END-SYSTOLIC AND END-DIASTOLIC CARDIAC MRI. THE TABLE COMPARES THE SYNTHMORPH (SM-SHAPES) AND VOXELMORPH (VM-NCC) MODELS PERFORMING BEST AT THIS TASK DESPITE OPTIMIZATION FOR BRAIN REGISTRATION, WITHOUT RETRAINING. A REDUCTION IN MSD TRANSLATES TO BETTER ALIGNMENT OF THE LEFT VENTRICULAR STRUCTURES. SD ABBREVIATES STANDARD DEVIATION, AND WE HIGHLIGHT THE BEST RESULT FOR EACH SET OF CONTOURS IN BOLD

| | Endocardium | | Epicardium | |
|---|---|---|---|---|
| | SynthMorph | VoxelMorph | SynthMorph | VoxelMorph |
| Pairs improved | 29/33 (88%) | 28/33 (85%) | **28/33 (85%)** | 25/33 (76%) |
| Mean ± SD (mm) | **−0.8 ± 0.1** | −0.7 ± 0.1 | **−0.6 ± 0.1** | −0.2 ± 0.1 |
| Mean ± SD (%) | **−10.2 ± 1.6** | −9.0 ± 1.9 | **−11.6 ± 1.5** | −4.3 ± 1.4 |
| Best pair (%) | **−35.4** | −33.8 | **−29.6** | −22.5 |
| Worst pair (%) | **+3.9** | +6.1 | **+6.0** | +13.2 |

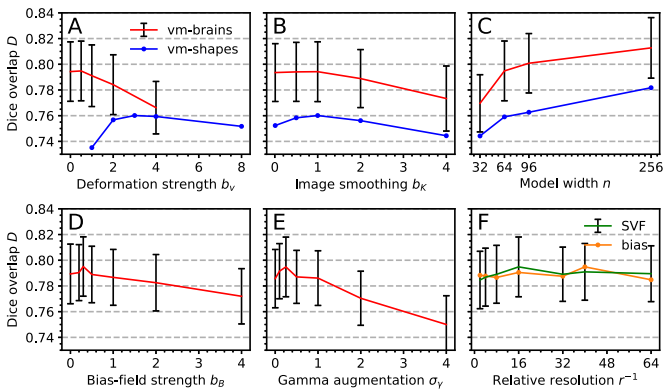

Fig. 13. Effect of training settings on median registration accuracy: (a) Maximum velocity-field SD $b_v$. (b) Maximum image-smoothing SD $b_K$. (c) Number of filters $n$ per convolutional layer. (d) Maximum bias-field SD $b_B$. (e) Gamma-augmentation SD $\sigma_\gamma$. (f) Resolution $r$. Error bars are comparable across methods and indicate SD over subjects.

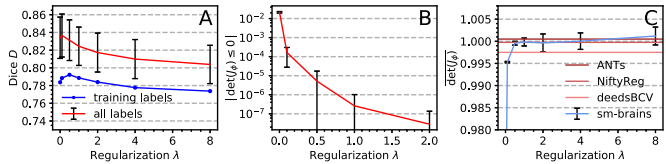

Fig. 14. Regularization analysis. (a) Median accuracy. Error bars are comparable across label sets and indicate SD over subjects. (b) Proportion of voxels where the warp $\phi$ folds, i.e. $\det(J_\phi) \leq 0$ for voxel Jacobian $J_\phi$ (0 for $\lambda > 1$; of $4.9 \times 10^6$ voxels). (c) Average Jacobian determinant. For $\lambda \geq 1$, the deviation from the ideal value 1 is less than $2 \times 10^{-3}$.

include smaller structures, the mean overlap $D$ reduces for $\lambda < 1$, as the network then focuses on optimizing the training structures. This does not apply to sm-shapes, which is agnostic to anatomy since we train it on all synthetic labels present in the random maps. Fig. 14b shows a small proportion of locations where the warp field folds, decreasing with increasing $\lambda$. For test results, we use $\lambda = 1$, where the proportion of folding voxels is below $10^{-6}$ at our numerical precision. At fixed $\lambda = 1$, increasing the number of integration steps reduces voxel folding, about 6-fold for 10 instead of 5 steps, after which further increases have no effect.

## H. Experiment 4: Cine-Cardiac Application

In this experiment we test SynthMorph and VoxelMorph on cine-cardiac MRI to assess how these models transfer to a domain with substantially different image content. The goal is to analyze whether already trained models extend beyond neuroimaging, rather than claiming their outperformance over methods specifically developed for the task. We choose the dataset because the trained networks assume affine registration of the input images, which can be challenging in non-brain applications, whereas cardiac frames from the same subject are largely aligned. This provides an opportunity for testing registration of images with structured background within contrast; we test cross-contrast registration in Section III-E and Section III-F.

*1) Setup:* Non-rigid registration of cardiac images from the same subject is an important tool that can help assess cardiovascular health. Some approaches choose an end-diastolic frame as the fixed image, as it is easily identified [95], [96]. Thus, we pair an end-systolic with an end-diastolic frame for each of 33 subjects, corresponding to maximum cardiac contraction and expansion. For 3D registration of these pairs, we use already trained SynthMorph and VoxelMorph models without optimizing for the new task.

*2) Results:* Table III compares the effect on mean symmetric surface distance for the best-performing SynthMorph (sm-shapes) and VoxelMorph (vm-ncc) models. Registration with sm-shapes reduces MSD between the epicardial contours by $\Delta S/S = (11.6 \pm 1.5)\%$ on average, improving MSD for 85% of pairs (lower MSD is better). The mean reduction for vm-ncc is only $\Delta S/S = (4.3 \pm 1.4)\%$. While the pairs that do not improve appear visually unchanged, MSD increases slightly: for example, the most substantial decrease for sm-shapes is 35.4%, but the least accurate registration only results in a 3.9% increase. While the performance gap between the models is smaller for endocardial MSD, sm-shapes still outperforms vm-ncc. The models sm-brains and vm-nmi underperform sm-shapes and vm-ncc in terms of MSD, respectively. Fig. 15 shows exemplary cardiac frames before and after registration with sm-shapes along with the displacement fields, illustrating how SynthMorph leaves most anatomy intact while focusing on dilation of the heart to match its late-diastolic shape.

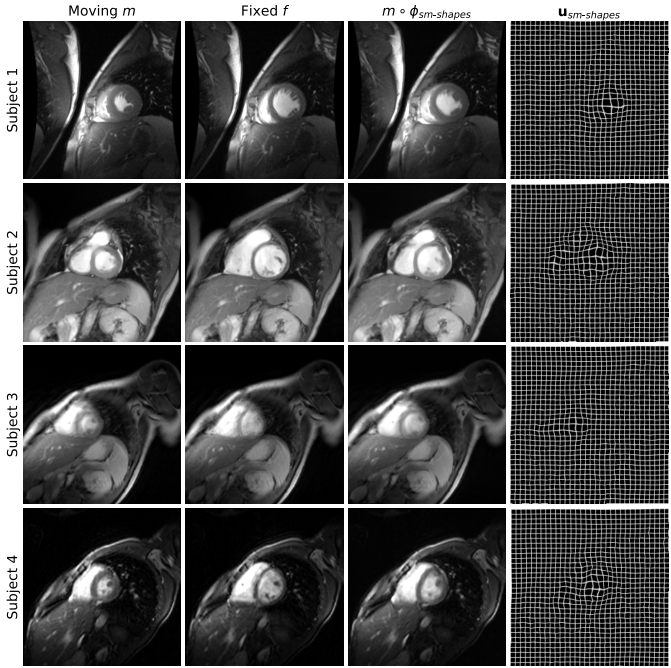

Fig. 15. Cine-cardiac registration results. Each row shows an image pair from a different subject: we register frames corresponding to maximum cardiac contraction and expansion, respectively. Despite the thick slices and more diverse image content than typical of neuroimaging data, `sm-shapes` clearly dilates the contracted anatomy as indicated by the displacement fields in the rightmost column.

## IV. DISCUSSION

We propose `SynthMorph`, a general strategy for learning contrast-invariant registration that does not require any imaging data during training. We remove the need for acquired data by synthesizing images randomly from noise distributions.

### A. Generalizability

A significant challenge in the deployment of neural networks is their generalizability to image types unseen during training. Existing learning methods like `VoxelMorph` achieve good registration performance but consistently fail for new MRI contrasts at test time. For example, `vm-ncc` is trained on T1w pairs and breaks down both across contrasts (e.g. T1w-T2w) *and* within new contrasts (e.g. T2w-T2w). The `SynthMorph` strategy addresses this weakness and makes networks resilient to contrast changes by exposing them to a wide range of synthetic images, far beyond the shapes and contrasts typical of MRI. This approach obviates the need for retraining to register images acquired with a new sequence.

Training conventional `VoxelMorph` with a loss evaluated on T1w images while augmenting the input contrasts enables the transfer of domain-specific specific knowledge to cross-contrast registration tasks. However, the associated decrease in within-contrast performance indicates the benefit of `SynthMorph`: learning to match anatomical features independent of their appearance in the gray-scale images.

The choice of optimum hyperparameters also is an important problem for many deep learning applications. While the grid search of Fig. 13 illustrates the dependency of accuracy on hyperparameter values, `SynthMorph` performance is robust over the ranges typical of medical imaging modalities, e.g. smoothing kernels with SD $\sigma_K \in [0, 2]$.

We select `SynthMorph` hyperparameters for all experiments based on the analysis of Fig. 13, using validation data that do not overlap with the test sets. The chosen parameters (Table I) enable robust registration across six different test sets in Section III-E and over a landscape of continually evolving MRI contrasts in Section III-F, demonstrating their generalizability across datasets.

### B. Baseline Comparison

Networks trained within the `SynthMorph` framework do not have access to the MRI contrasts of the test set nor indeed to any MRI data at all. Yet `sm-shapes` matches state-of-the-art classical performance within contrasts and provides substantial improvements in cross-contrast performance over `ANTs` and `NiftyReg`, all while being substantially faster.

Registration accuracy varies with the particular contrast pairings, likely because anatomical structures appear different on images acquired with different MRI sequences. There is no guarantee that a structure will have contrast with neighboring structures and can be registered well to a scan of a particular MRI contrast (e.g. PDw). Nevertheless, `SynthMorph` outperforms both classical and learning-based methods across contrasts, demonstrating that it can indeed register new image types, to the extent permitted by the intrinsic contrast.

If brain segmentations are available, including these in the image synthesis enables the `sm-brains` network to outperform all methods tested by a substantial margin–at any contrast combination tested–although this model still does not require any acquired MR images during training.

Visual inspection of typical deformation fields in Fig. 6 provides an interesting insight: the `sm-brains` network appears to learn to identify the structures of interest optimized in the loss. Thus, it focuses on registering these brain regions and their close neighbors, while leaving the background and structures such as the skull unaffected. This anatomical knowledge enables registration of skull-stripped images to data including the full head. While the resulting deformations may appear less regular than those estimated by classical methods, our analysis of the Jacobian determinant demonstrates comparable field regularity across methods.

### C. Dice-Loss Sensitivity

When training on synthesized structures with arbitrary geometry, the network learns to generally match shapes based on contrast. The `sm-shapes` model does not learn to register specific human anatomical structures or sub-structures since we never expose it to specific neuroanatomy and instead sample random shapes of all sizes during training. In the experiment trained on brain anatomy, the model matches substructures within labels if they manifest contrast. If substructures are not discernible, the smooth regularization yields reasonable predictions. This can be observed with `sm-brains` for smaller structures that are not included in the dissimilarity loss $\mathcal{L}_{dis}$ but for which we obtain competitive validation Dice scores, e.g. the 3rd and 4th ventricle.

### D. Supervised or Unsupervised?

Since ground-truth deformation fields are available for `sm-shapes`, we also train baseline models in a supervised manner. This approach consistently under-performs its unsupervised counterpart, for which we propose three possible explanations. First, several different deformations can result in the same warped brain, which has the potential to introduce a level of ambiguity into the registration problem that makes it challenging to train a reliable predictor. Second, related to this point, image areas with little intensity variation such as the background or central parts of the white matter offer no guidance for the supervised network to match the arbitrary ground-truth deformation, compared to unsupervised models, that are driven by the regularization term in those areas. Third, the synthesized transforms may not represent an exact identifiable mapping between the source and target image because of errors introduced by nearest-neighbor interpolation of the input label maps and further augmentation steps including image blurring and additive noise.

### E. Further Work

While `SynthMorph` addresses important drawbacks of within and between-contrast registration methods, it can be expanded in several ways.

First, we plan to extend our framework to incorporate affine registration [47], [54], [55], [97]. We will explore whether the simultaneous estimation of affine and deformable transforms can improve accuracy and thoroughly investigate the appropriateness of architectures for doing this in heterogeneous data. In the current work, the input images $\{m, f\}$ need prior affine alignment for optimal results. Although this preprocessing step is beyond the focus of our current contribution, the code we make available includes an optimization-based affine solution, thus providing full registration capabilities independent of third-party tools. The optimization estimates 12 affine parameters for each new pair of 3D images in ~10 seconds, with accuracy comparable to `ANTs` and `NiftyReg`.

Second, our approach promises to be extensible to unprocessed images acquired with any MRI sequence, of any body part, possibly even beyond medical imaging. While this is an exciting area of research, the present work focuses on neuroimaging applications since the breadth of the analyses required is beyond the scope of a single solid contribution.

Third, an obvious extension is to combine the simulation strategy with existing image data that might already be available. We plan to investigate whether including real MRI scans would aid, or instead bias the network and reduce its ability to generalize to unseen contrast variations.

### F. Invariant Representations

We investigate why the `SynthMorph` strategy enables substantial improvements in registration performance. In particular, we evaluate how accuracy responds to gradual changes in MRI contrast and show that the deep layers of `SynthMorph` models exhibit a greater degree of invariance to contrast changes than networks trained in a conventional fashion. We present qualitative and quantitative analyses demonstrating that the enhanced contrast invariance leads to highly robust registration across wide spectra of MR images simulated for two commonly used pulse sequences, FLASH and MPRAGE.

### G. Cardiac Registration

The cine-cardiac experiment demonstrates the viability and potential of `SynthMorph` applied to a domain with substantially different image content than neuroimaging. While we do not claim to outperform dedicated cardiac registration methods, `sm-shapes` reduces the MSD metric between the fixed and moving frames in the majority of subjects, to a greater extent than any of the `sm-brains`, `vm-ncc`, and `vm-nmi` models. The network achieves this result without any optimization for the anatomy or image type considered, using weights obtained with generation hyperparameters tuned for isotropic 3D brain registration. In contrast, the cardiac data are volumes resampled from stacks of slices with thicknesses exceeding the voxel dimension of our neuroimaging test sets by 9-fold on average. Although `sm-shapes` is not an optimized registration tool for cardiac MRI, its weights provide a great choice for initializing networks when training application-specific registration, since the model produces reasonable results and is unbiased towards any particular anatomy.

### H. Domain-Specific Knowledge

The comparisons between `sm-brains` and `sm-shapes` in neuroimaging datasets indicate that `SynthMorph` performs substantially better when exploiting domain-specific knowledge. For the cardiac application, this could be achieved in the following ways. First, if the amplitude of cardiac motion exceeds the deformations sampled during `sm-shapes` training, increasing hyperparameter $b_v$ will be beneficial. Second, a lower regularization weight $\lambda$ may be favorable for cardiac motion, which is characterized by considerable displacements within a small portion of space. Third, anatomical segmentations in fields other than neuroimaging often include fewer different labels. To overcome this challenge and synthesize images complex enough for networks to learn anatomy-specific registration, these label maps could be augmented by including arbitrary geometric shapes as diverse backgrounds.

Qualitatively, our experience is that generation hyperparameters represent a trade-off between (1) sampling from a distribution large enough to include the features of a target dataset while promoting network robustness by exposure to broad variability, and (2) ensuring that the network capacity is adequate for capturing the sampled variation. As an alternative to making domain-specific informed changes to the generation hyperparameters and retraining networks, recent work suggests to optimize hyperparameter values efficiently at test time using hypernetworks [98]. In addition to a registration pair, such hypernetworks take as input a set of hyperparameters and output the weights of a registration network, thus modeling a continuum of registration networks each trained with different hyperparameter values.

### I. Data Requirements for Registration

The baseline comparison reveals that neither augmenting nor adding data in `VoxelMorph` training boosts performance. While counter-intuitive to intuitions about deep learning in classification tasks, this result is consistent with recent findings confirming that large datasets are not necessary for tasks like deformable registration and segmentation, that have sizable input and output spaces [58], [59], [99]: in effect, every image voxel can be thought of as a data sample, although these are, of course, not independent. For example, reasonable segmentation performance can be achieved with only a handful of annotated images [58]. For registration, our analysis shows that `SynthMorph` training with label maps from only 40 subjects enables outperformance of all other methods tested.

We train the `VoxelMorph` baseline using images from 100 subjects, randomly flipping the axes of each input pair, which already gives rise to 79,200 different cross-subject image combinations. An analysis in the `VoxelMorph` paper [17] comparing training sets of size 100 and 3231 without randomly flipping axes provides further evidence that larger datasets do not necessarily lead to significant performance gains.

## V. Conclusion

Our study establishes the utility of training on synthetic data only and indicates a novel way of thinking about feature invariance in the context of registration. `SynthMorph` enables users to build on the strengths of deep learning, including rapid execution, increased robustness to local minima and outliers, and flexibility in the choice of loss functions, by now having the previously-missing ability to generalize to any MRI contrast at test time. This leads us to believe the strategy can be broadly applied to networks to limit the need for training data while vastly improving applicability.

## Acknowledgment

The authors thank Danielle F. Pace for help with surface distances. Data are provided in part by OASIS Cross-Sectional (PIs D. Marcus, R. Buckner, J. Csernansky, J. Morris; NIH grants P50 AG05681, P01 AG03991, P01 AG026276, R01 AG021910, P20 MH071616, U24 R021382). HCP-A: Research reported in this publication is supported by Grant U01 AG052564 and Grant AG052564-S1 and by the 14 NIH Institutes and Centers that support the NIH Blueprint for Neuroscience Research, by the McDonnell Center for Systems Neuroscience at Washington University, by the Office of the Provost at Washington University, and by the University of Minnesota Medical School.

## References

[1] D. W. McRobbie, E. A. Moore, M. J. Graves, and M. R. Prince, *MRI From Picture to Proton*. Cambridge, U.K.: Cambridge Univ. Press, 2017.

[2] M. Hoffmann, T. A. Carpenter, G. B. Williams, and S. J. Sawiak, "A survey of patient motion in disorders of consciousness and optimization of its retrospective correction," *Magn. Reson. Imag.*, vol. 33, no. 3, pp. 346–350, Apr. 2015.

[3] J. Hajnal and D. Hill, *Medical Image Registration* (Biomedical Engineering). Boca Raton, FL, USA: CRC Press, 2001.

[4] B. Fischl, "Freesurfer," *NeuroImage*, vol. 62, no. 2, pp. 774–781, 2012.

[5] R. S. Frackowiak, *Human Brain Function*. Amsterdam, The Netherlands: Elsevier, 2004.

[6] O. Puonti, J. E. Iglesias, and K. V. Leemput, "Fast and sequence-adaptive whole-brain segmentation using parametric Bayesian modeling," *NeuroImage*, vol. 143, pp. 235–249, Dec. 2016.

[7] R. Sridharan *et al.*, "Quantification and analysis of large multimodal clinical image studies: Application to stroke," in *Multimodal Brain Image Analysis*. Cham, Switzerland: Springer, 2013, pp. 18–30.

[8] M. Goubran *et al.*, "Multimodal image registration and connectivity analysis for integration of connectomic data from microscopy to MRI," *Nature Commun.*, vol. 10, no. 1, pp. 1–17, Dec. 2019.

[9] B. C. Lee, M. K. Lin, Y. Fu, J. Hata, M. I. Miller, and P. P. Mitra, "Multimodal cross-registration and quantification of metric distortions in marmoset whole brain histology using diffeomorphic mappings," *J. Comput. Neurol.*, vol. 529, no. 2, pp. 281–295, 2020.

[10] M. Lorenzi, N. Ayache, G. Frisoni, and X. Pennec, "LCC-Demons: A robust and accurate symmetric diffeomorphic registration algorithm," *NeuroImage*, vol. 81, pp. 470–483, Nov. 2013.

[11] J. Ashburner, "A fast diffeomorphic image registration algorithm," *NeuroImage*, vol. 38, no. 1, pp. 113–195, 2007.

[12] B. B. Avants, C. L. Epstein, M. Grossman, and J. C. Gee, "Symmetric diffeomorphic image registration with cross-correlation: Evaluating automated labeling of elderly and neurodegenerative brain," *Med. Image Anal.*, vol. 12, no. 1, pp. 26–41, 2008.

[13] M. Modat *et al.*, "Fast free-form deformation using graphics processing units," *Comput. Meth. Prog. Biomed.*, vol. 98, no. 3, pp. 278–284, 2010.

[14] K. Rohr, H. S. Stiehl, R. Sprengel, T. M. Buzug, J. Weese, and M. Kuhn, "Landmark-based elastic registration using approximating thin-plate splines," *IEEE Trans. Med. Imag.*, vol. 20, no. 6, pp. 526–534, Jun. 2001.

[15] D. Rueckert, L. I. Sonoda, C. Hayes, D. L. Hill, M. O. Leach, and D. J. Hawkes, "Nonrigid registration using free-form deformations: Application to breast mr images," *IEEE Trans. Med. Imag.*, vol. 18, no. 8, pp. 712–721, Aug. 1999.

[16] T. Vercauteren, X. Pennec, A. Perchant, and N. Ayache, "Diffeomorphic demons: Efficient non-parametric image registration," *NeuroImage*, vol. 45, no. 1, pp. S61–S72, Mar. 2009.

[17] G. Balakrishnan, A. Zhao, M. Sabuncu, J. Guttag, and A. V. Dalca, "Voxelmorph: A learning framework for deformable medical image registration," *IEEE Trans. Med. Imag.*, vol. 38, no. 8, pp. 1788–1800, Feb. 2019.

[18] B. D. de Vos, F. F. Berendsen, M. A. Viergever, M. Staring, and I. Išgum, "End-to-end unsupervised deformable image registration with a convolutional neural network," in *Deep Learning in Medical Image Analysis and Multimodal Learning for Clinical Decision*. Cham, Switzerland: Springer, 2017, pp. 204–212.

[19] C. Guetter, C. Xu, F. Sauer, and J. Hornegger, "Learning based non-rigid multi-modal image registration using kullback-leibler divergence," in *Medical Image Computing and Computer-Assisted Intervention*. Berlin, Germany: Springer, 2005, pp. 255–262.

[20] H. Li and Y. Fan, "Non-rigid image registration using fully convolutional networks with deep self-supervision," 2017, *arXiv:1709.00799*. [Online]. Available: http://arxiv.org/abs/1709.00799

[21] M. Rohé, M. Datar, T. Heimann, M. Sermesant, and X. Pennec, "SVF-Net: Learning deformable image registration using shape matching," in *Proc. Int. Conf. Medical Image Comput. Comput.-Assist. Intervent.*, 2017, pp. 266–274.

[22] H. Sokooti, B. de Vos, F. Berendsen, B. P. Lelieveldt, I. Išgum, and M. Staring, "Nonrigid image registration using multi-scale 3D convolutional neural networks," in *Medical Image Computing and Computer Assisted Intervention*. Cham, Switzerland: Springer, 2017, pp. 232–239.

[23] G. Wu, M. Kim, Q. Wang, B. C. Munsell, and D. Shen, "Scalable high-performance image registration framework by unsupervised deep feature representations learning," *IEEE Trans. Biomed. Eng.*, vol. 63, no. 7, pp. 1505–1516, Jul. 2016.

[24] X. Yang, R. Kwitt, M. Styner, and M. Niethammer, "Quicksilver: Fast predictive image registration—A deep learning approach," *NeuroImage*, vol. 158, pp. 378–396, Jun. 2017.

[25] M. F. Beg, M. I. Miller, A. Trouvé, and L. Younes, "Computing large deformation metric mappings via geodesic flows of diffeomorphisms," *Int. J. Comput. Vis.*, vol. 61, no. 2, pp. 139–157, 2005.

[26] A. Klein *et al.*, "Evaluation of 14 nonlinear deformation algorithms applied to human brain MRI registration," *NeuroImage*, vol. 46, no. 3, pp. 786–802, 2009.

[27] P. Viola and W. M. Wells III, "Alignment by maximization of mutual information," *Int. J. Comput. Vis.*, vol. 24, no. 2, pp. 137–154, 1997.

[28] A. Roche, G. Malandain, X. Pennec, and N. Ayache, "The correlation ratio as a new similarity measure for multimodal image registration," in *Proc. Int. Conf. Med. Image Comput. Comput.-Assist. Intervent.*, 1998, pp. 1115–1124.

[29] J. E. Iglesias, E. Konukoglu, D. Zikic, B. Glocker, K. Van Leemput, and B. Fischl, "Is synthesizing MRI contrast useful for inter-modality analysis," in *Medical Image Computing and Computer-Assisted Intervention*. Berlin, Germany: Springer, 2013, pp. 631–638.

[30] Y. Xiao *et al.*, "Evaluation of MRI to ultrasound registration methods for brain shift correction: The CuRIOUS2018 challenge," *IEEE Trans. Med. Imag.*, vol. 39, no. 3, pp. 777–786, Mar. 2020.

[31] B. van Ginneken, S. Kerkstra, and J. Meakin. (2019). *Medical Image Computing and Computer-Assisted Intervention Curious*. [Online]. Available: https://curious2019.grand-challenge.org

[32] E. Haber and J. Modersitzki, "Intensity gradient based registration and fusion of multi-modal images," in *Medical Image Computing and Computer-Assisted Intervention*. Berlin, Germany: Springer, 2006, pp. 726–733.

[33] M. P. Heinrich *et al.*, "MIND: Modality independent neighbourhood descriptor for multi-modal deformable registration," *Med. Image. Anal.*, vol. 16, no. 7, pp. 1423–1435, 2012.

[34] M. Mellor and M. Brady, "Phase mutual information as a similarity measure for registration," *Med. Image. Anal.*, vol. 9, no. 4, pp. 330–343, 2005.

[35] C. Wachinger and N. Navab, "Entropy and Laplacian images: Structural representations for multi-modal registration," *Med. Image Anal.*, vol. 16, no. 1, pp. 1–17, Jan. 2012.

[36] Z. Xu *et al.*, "Evaluation of six registration methods for the human abdomen on clinically acquired CT," *IEEE Trans. Biomed. Eng.*, vol. 63, no. 8, pp. 1563–1572, Aug. 2016.

[37] L. König and J. Rühaak, "A fast and accurate parallel algorithm for non-linear image registration using normalized gradient fields," in *Proc. ISBI*, 2014, pp. 580–583.

[38] L. Konig, A. Derksen, M. Hallmann, and N. Papenberg, "Parallel and memory efficient multimodal image registration for radiotherapy using normalized gradient fields," in *Proc. IEEE 12th Int. Symp. Biomed. Imag. (ISBI)*, Apr. 2015, pp. 734–738.

[39] J. Rühaak *et al.*, "A fully parallel algorithm for multimodal image registration using normalized gradient fields," in *Proc. ISBI*, 2013, pp. 572–575.

[40] F. Kanavati *et al.*, "Supervoxel classification forests for estimating pairwise image correspondences," *Pattern Recognit.*, vol. 63, pp. 561–569, Mar. 2017.

[41] M. P. Heinrich, I. J. A. Simpson, B. W. Papież, M. Brady, and J. A. Schnabel, "Deformable image registration by combining uncertainty estimates from supervoxel belief propagation," *Med. Image Anal.*, vol. 27, pp. 57–71, Jan. 2016.

[42] K. A. J. Eppenhof and J. P. W. Pluim, "Pulmonary CT registration through supervised learning with convolutional neural networks," *IEEE Trans. Med. Imag.*, vol. 38, no. 5, pp. 1097–1105, May 2019.

[43] J. Krebs *et al.*, "Robust non-rigid registration through agent-based action learning," in *Proc. Int. Conf. Med. Image Comput. Comput.-Assist. Intervent.*, 2017, pp. 344–352.

[44] X. Yang, R. Kwitt, M. Styner, and M. Niethammer, "Fast predictive multimodal image registration," in *Proc. IEEE 14th Int. Symp. Biomed. Imag. (ISBI)*, Apr. 2017, pp. 48–57.

[45] A. V. Dalca, G. Balakrishnan, J. Guttag, and M. Sabuncu, "Unsupervised learning of probabilistic diffeomorphic registration for images and surfaces," *Med. Image. Anal.*, vol. 57, pp. 226–236, Oct. 2019.

[46] J. Krebs, H. Delingette, B. Mailhé, N. Ayache, and T. Mansi, "Learning a probabilistic model for diffeomorphic registration," *IEEE Trans. Med. Imag.*, vol. 38, no. 9, pp. 2165–2176, Sep. 2019.

[47] B. D. de Vos, F. F. Berendsen, M. A. Viergever, H. Sokooti, M. Staring, and I. Išgum, "A deep learning framework for unsupervised affine and deformable image registration," *Med. Image. Anal.*, vol. 52, pp. 128–143, Feb. 2019.

[48] C. K. Guo, "Multi-modal image registration with unsupervised deep learning," Ph.D. dissertation, Dept. Elect. Eng. Comput. Sci., Massachusetts Inst. Technol., Cambridge, MA, USA, 2019.

[49] M. Chen, A. Carass, A. Jog, J. Lee, S. Roy, and J. L. Prince, "Cross contrast multi-channel image registration using image synthesis for MR brain images," *Med. Image Anal.*, vol. 36, pp. 2–14, Feb. 2017.

[50] S. Roy, A. Carass, A. Jog, J. L. Prince, and J. Lee, "MR to CT registration of brains using image synthesis," *Proc. SPIE Med. Imag. Process.*, vol. 9034, Mar. 2014, Art. no. 903419.

[51] C. Bhushan, J. P. Haldar, S. Choi, A. A. Joshi, D. W. Shattuck, and R. M. Leahy, "Co-registration and distortion correction of diffusion and anatomical images based on inverse contrast normalization," *NeuroImage*, vol. 115, pp. 269–280, Jul. 2015.

[52] C. Tanner, F. Ozdemir, R. Profanter, V. Vishnevsky, E. Konukoglu, and O. Goksel, "Generative adversarial networks for MR-CT deformable image registration," 2018, *arXiv:1807.07349*. [Online]. Available: https://arxiv.org/abs/1807.07349

[53] C. Qin, B. Shi, R. Liao, T. Mansi, D. Rueckert, and A. Kamen, "Unsupervised deformable registration for multi-modal images via disentangled representations," in *Proc. IPMI*. Cham, Switzerland: Springer, 2019, pp. 249–261.

[54] Y. Hu *et al.*, "Weakly-supervised convolutional neural networks for multimodal image registration," *Med. Image Anal.*, vol. 49, pp. 1–13, Oct. 2018.

[55] Y. Hu *et al.*, "Label-driven weakly-supervised learning for multimodal deformable image registration," in *Proc. IEEE 15th Int. Symp. Biomed. Imag. (ISBI)*, Apr. 2018, pp. 1070–1074.

[56] A. Hering, S. Kuckertz, S. Heldmann, and M. P. Heinrich, "Enhancing label-driven deep deformable image registration with local distance metrics for state-of-the-art cardiac motion tracking," in *Bildverarbeitung Medizin*. Wiesbaden, Germany: Springer Vieweg, 2019, pp. 309–314.

[57] L. Mansilla, D. H. Milone, and E. Ferrante, "Learning deformable registration of medical images with anatomical constraints," *Neural Netw.*, vol. 124, pp. 269–279, Oct. 2020.

[58] H. W. Lee, M. R. Sabuncu, and A. V. Dalca, "Few labeled atlases are necessary for deep-learning-based segmentation," in *Proc. Mach. Learn. Health*, 2019, pp. 1–9.

[59] K. Chaitanya, N. Karani, C. F. Baumgartner, A. Becker, O. Donati, and E. Konukoglu, "Semi-supervised and task-driven data augmentation," in *Proc. Int. Conf. Inf. Process. Med. Imag.* Cham, Switzerland: Springer, 2019, pp. 29–41.

[60] O. Ronneberger, P. Fischer, and T. Brox, "U-net: Convolutional networks for biomedical image segmentation," *CoRR*, vol. abs/1505.04597, pp. 1–8, May 2015.

[61] J. Xu *et al.*, "Fetal pose estimation in volumetric MRI using a 3D convolution neural network," in *Medical Image Computing and Computer-Assisted Intervention*. Cham, Switzerland: Springer, 2019, pp. 403–410.

[62] A. Zhao, G. Balakrishnan, F. Durand, J. V. Guttag, and A. V. Dalca, "Data augmentation using learned transforms for one-shot medical image segmentation," *CoRR*, vol. abs/1902.09383, pp. 1–4, Feb. 2019.

[63] K. Kamnitsas *et al.*, "Unsupervised domain adaptation in brain lesion segmentation with adversarial networks," in *Proc. IPMI*, 2017, pp. 597–609.

[64] B. Billot, D. Greve, K. Van Leemput, B. Fischl, J. E. Iglesias, and A. V. Dalca, "A learning strategy for contrast-agnostic mri segmentation," in *Proc. PMLR*, vol. 121, Montreal, QC, Canada, Jul. 2020, pp. 75–93.

[65] M. Hoffmann, B. Billot, J. E. Iglesias, B. Fischl, and A. V. Dalca, "Learning MRI contrast-agnostic registration," in *Proc. ISBI*, 2021, pp. 899–903.

[66] A. V. Dalca. (2018). *VoxelMorph: Learning-Based Image Registration*. [Online]. Available: https://voxelmorph.net

[67] F. Milletari, N. Navab, and S.-A. Ahmadi, "V-net: Fully convolutional neural networks for volumetric medical image segmentation," in *Proc. 3DV*, 2016, pp. 565–571.

[68] V. Arsigny, O. Commowick, X. Pennec, and N. Ayache, "A log-euclidean framework for statistics on diffeomorphisms," in *Medical Image Computing and Computer-Assisted Intervention*. Berlin, Germany: Springer, 2006, pp. 924–931.

[69] J. Ashburner and K. J. Friston, "Unified segmentation," *NeuroImage*, vol. 26, pp. 839–851, Oct. 2005.

[70] K. Van Leemput, F. Maes, D. Vandermeulen, and P. Suetens, "Automated model-based tissue classification of MR images of the brain," *IEEE Trans. Med. Imag.*, vol. 18, no. 10, pp. 897–908, Oct. 1999.

[71] W. M. Wells, W. E. L. Grimson, R. Kikinis, and F. A. Jolesz, "Adaptive segmentation of MRI data," *IEEE Trans. Med. Imag.*, vol. 15, no. 4, pp. 429–442, Aug. 1996.

[72] Y. Zhang, M. Brady, and S. Smith, "Segmentation of brain MR images through a hidden Markov random field model and the expectation-maximization algorithm," *IEEE Trans. Med. Imag.*, vol. 20, no. 1, pp. 45–57, Jan. 2001.

[73] K. Van Leemput, F. Maes, D. Vandermeulen, and P. Suetens, "A unifying framework for partial volume segmentation of brain MR images," *IEEE Trans. Med. Imag.*, vol. 22, no. 1, pp. 105–119, Apr. 2003.

[74] B. Belaroussi, J. Milles, S. Carme, Y. M. Zhu, and H. Benoit-Cattin, "Intensity non-uniformity correction in MRI: Existing methods and their validation," *Med. Image. Anal.*, vol. 10, no. 2, pp. 234–246, 2006.

[75] Z. Hou, "A review on MR image intensity inhomogeneity correction," *Int. J. Biomed. Imag.*, vol. 2006, pp. 1–11, Oct. 2006.

[76] M. Abadi *et al.*, "Tensorflow: A system for large-scale machine learning," in *Proc. USENIX Symp. Oper. Syst. Design Implement.*, 2016, pp. 265–283.

[77] D. P. Kingma and J. Ba, "Adam: A method for stochastic optimization," 2014, *arXiv:1412.6980*. [Online]. Available: http://arxiv.org/abs/1412.6980

[78] D. S. Marcus, T. H. Wang, J. Parker, J. G. Csernansky, J. C. Morris, and R. L. Buckner, "Open Access Series Of Imaging Studies (OASIS): Cross-sectional MRI data in young, middle aged, nondemented, and demented older adults," *J. Cogn. Neurosci.*, vol. 19, no. 9, pp. 1498–1507, 2007.

[79] S. Y. Bookheimer *et al.*, "The lifespan human connectome project in aging: An overview," *NeuroImage*, vol. 185, pp. 335–348, Oct. 2019.

[80] M. P. Harms *et al.*, "Extending the human connectome project across ages: Imaging protocols for the lifespan development and aging projects," *NeuroImage*, vol. 183, pp. 972–984, Dec. 2018.

[81] A. J. W. van der Kouwe, T. Benner, D. H. Salat, and B. Fischl, "Brain morphometry with multiecho MPRAGE," *NeuroImage*, vol. 40, no. 2, pp. 559–569, Apr. 2008.

[82] J. P. Mugler, "Optimized three-dimensional fast-spin-echo MRI," *J. Magn. Reson. Imag.*, vol. 39, no. 4, pp. 745–767, Apr. 2014.

[83] D. B. Keator *et al.*, "A national human neuroimaging collaboratory enabled by the biomedical informatics research network (BIRN)," *IEEE Trans. Inf. Technol. Biomed.*, vol. 12, no. 2, pp. 162–172, Mar. 2008.

[84] C. Sudlow *et al.*, "UK biobank: An open access resource for identifying the causes of a wide range of complex diseases of middle and old age," *PLOS Med.*, vol. 12, no. 3, Mar. 2015, Art. no. e1001779.

[85] A. J. Holmes *et al.*, "Brain genomics superstruct project initial data release with structural, functional, and behavioral measures," *Sci. Data*, vol. 2, no. 1, Dec. 2015, Art. no. 150031.

[86] R. B. Buxton, R. R. Edelman, B. R. Rosen, G. L. Wismer, and T. J. Brady, "Contrast in rapid mr imaging: T1- and t2-weighted imaging," *J Comput Assist Tomogr*, vol. 11, no. 1, pp. 7–16, 1987.

[87] J. P. Marques, T. Kober, G. Krueger, W. van der Zwaag, P.-F. Van de Moortele, and R. Gruetter, "MP2RAGE, a self bias-field corrected sequence for improved segmentation and T1-mapping at high field," *NeuroImage*, vol. 49, no. 2, pp. 1271–1281, 2010.

[88] B. Fischl *et al.*, "Whole brain segmentation: Automated labeling of neuroanatomical structures in the human brain," *Neuron*, vol. 33, no. 3, pp. 341–355, 2002.

[89] J. D. Van Horn *et al.*, "The functional magnetic resonance imaging data center (fMRIDC): The challenges and rewards of large–scale databasing of neuroimaging studies," *Phil. Trans. Roy. Soc. London. Ser. B, Biol. Sci.*, vol. 356, no. 1412, pp. 1323–1339, Aug. 2001.

[90] A. Andreopoulos and J. K. Tsotsos, "Efficient and generalizable statistical models of shape and appearance for analysis of cardiac MRI," *Med. Image Anal.*, vol. 12, no. 3, pp. 335–357, 2008.

[91] M. Reuter, H. D. Rosas, and B. Fischl, "Highly accurate inverse consistent registration: A robust approach," *NeuroImage*, vol. 53, no. 4, pp. 1181–1196, 2010.

[92] D. Pustina and P. Cook. (2017). *Anatomy of a Registration Call*. [Online]. Available: https://github.com/ANTsX/ANTs/wiki/Anatomy-of-an-antsRegistration-call

[93] M. P. Heinrich, M. Jenkinson, M. Brady, and J. A. Schnabel, "MRF-based deformable registration and ventilation estimation of lung CT," *IEEE Trans. Med. Imag.*, vol. 32, no. 7, pp. 1239–1248, Jul. 2013.

[94] L. R. Dice, "Measures of the amount of ecologic association between species," *Ecology*, vol. 26, no. 3, pp. 297–302, 1945.

[95] W. Shi *et al.*, "A comprehensive cardiac motion estimation framework using both untagged and 3-D tagged MR images based on nonrigid registration," *IEEE Trans. Med. Imag.*, vol. 31, no. 6, pp. 1263–1275, Jun. 2012.

[96] M. J. Ledesma-Carbayo *et al.*, "Spatio-temporal nonrigid registration for ultrasound cardiac motion estimation," *IEEE Trans. Med. Imag.*, vol. 24, no. 9, pp. 1113–1126, Sep. 2005.

[97] E. Chee and Z. Wu, "AIRNet: Self-supervised affine registration for 3D medical images using neural networks," 2018, *arXiv:1810.02583*. [Online]. Available: http://arxiv.org/abs/1810.02583

[98] A. Hoopes, M. Hoffmann, B. Fischl, J. Guttag, and A. V. Dalca, "Hypermorph: Amortized hyperparameter learning for image registration," in *Proc. Int. Conf. Inf. Process. Med. Imag.* Cham, Switzerland: Springer, 2021, pp. 3–17.

[99] A. Zhao, G. Balakrishnan, F. Durand, J. V. Guttag, and A. V. Dalca, "Data augmentation using learned transformations for one-shot medical image segmentation," in *Proc. IEEE/CVF Conf. Comput. Vis. Pattern Recognit.*, Oct. 2019, pp. 8543–8553.